\documentclass[12pt,oneside,reqno]{amsart}

\usepackage[flushleft]{caption}
\usepackage{color}
\usepackage{subfig}
\usepackage{graphicx}
\usepackage{verbatim}
\usepackage{comment}
\usepackage{calc}
\usepackage{soul}
\usepackage{geometry}
\usepackage{marginnote} %%% for margin notes

\usepackage{setspace}
\usepackage{amsthm}

\newtheorem{theorem}{Theorem}[section]

\theoremstyle{definition}

\theoremstyle{remark}

\allowdisplaybreaks

\begin{document}
\thispagestyle{empty}

\renewcommand\rightmark{Interpretation and inference for altmetric indicators}%%%Running head%%%
\renewcommand\leftmark{Interpretation and inference for altmetric indicators}%%%Running head%%%

\begin{center}
\textbf{Interpretation and inference for altmetric indicators arising\\ from sparse data statistics}
\end{center}

\vspace {0.3 cm}

\noindent
{\tiny Dedicated to the memory of Brian Marx and the many meetings we had discussing this article.  He was a warmhearted and modest colleague and a distinguished statistician, who was a Fellow of the American Statistical Association.}

\vspace {0.3 cm}

\begin{center}
by

Lawrence Smolinsky,${}^1$ Bernhard Klingenberg,${}^{2,3}$ and Brian D. Marx${}^4$
\end{center}

\vspace {0.3 cm}

\noindent
${}^1${\footnotesize{Department of Mathematics, Louisiana State University, Baton Rouge, LA, USA}} \vspace{-0.3cm} \\ \\
${}^2${\footnotesize{Department of Mathematics and Statistics, Williams College, Williamstown, MA, USA}}  \vspace{-0.3cm} \\ \\
${}^3${\footnotesize{Graduate Program in Data Science, New College of Florida, Sarasota, FL, USA}}  \vspace{-0.3cm} \\ \\
${}^4${\footnotesize{Department of Experimental Statistics, Louisiana State University, Baton Rouge, LA, USA}}

\vspace {0.5 cm}

\noindent
\textbf{Highlights}
% 5 bullet points (Each with maximum 85 characters, including spaces)

\noindent
$\bullet$ We study the altmetric indicator MHq and introduce a new indicator, MHRR.\\
$\bullet$ The published interpretation and confidence intervals of MHq are shown to be wrong.\\
$\bullet$ We give a correct interpretation and correct confidence interval for MHq.\\
$\bullet$ MHRR is a weighted sum of risk ratios and is more natural and intuitive than MHq.\\
$\bullet$ A computer simulation study validates and compares confidence interval formulas.

\vspace {0.5 cm}

% \vfil
% \pagebreak

\begin{center} \textbf{Abstract}
\end{center}
%In this article, we explore the meaning and properties of altmetric indicators, while correcting some misconceptions and errors in recent literature. In 2018 Bornmann and Haunschild (2018a) introduced a new indicator called the Mantel-Haenszel quotient (MHq) to measure alternative metrics (or altmetrics) of scientometric data.  Due to errors in the original literature, MHq is not what it originally seemed.  First, conditional probabilities were misunderstood which led to an incorrect interpretation.  Nevertheless, MHq is a meaningful statistic, and we give its interpretation.  Second, a variance estimator was incorrectly derived leading to incorrect confidence intervals.  We derive a new variance estimator for MHq, which leads to narrower confidence intervals.  Simulations show the improved performance of our estimator and confidence interval to the ones originally proposed.  Since MHq does not match its original description in the literature, we propose a new indicator, the Mantel-Haenszel row risk ratio (MHRR), to meet that need.  For both MHRR and MHq, a value greater (less) than one means performance is better (worse) than the reference set called the world.  The indicator MHq will require large samples of articles or documents, but MHRR does not.

In 2018 Bornmann and Haunschild (2018a) introduced a new indicator called the Mantel-Haenszel quotient (MHq) to measure alternative metrics (or altmetrics) of scientometric data.  In this article we review the Mantel-Haenszel statistics, point out two errors in the literature, and introduce a new indicator.  First, we correct the interpretation of MHq and mention that it is still a meaningful indicator. Second, we correct the variance formula for MHq, which leads to narrower confidence intervals.  A simulation study shows the superior performance of our variance estimator and confidence intervals. Since MHq does not match its original description in the literature, we propose a new indicator, the Mantel-Haenszel row risk ratio (MHRR), to meet that need.  Interpretation and statistical inference for MHRR are discussed.  For both MHRR and MHq, a value greater (less) than one means performance is better (worse) than in the reference set called the world.

\textit{Keywords}:  Bibliometrics, Altmetrics, Mantel-Haenszel quotient (MHq), Mantel-Haenszel row risk ratio (MHRR), relative risk, risk ratio

\vfil
\pagebreak

\section {\textbf{Introduction}}

The study of scientific and academic publication output assesses the impact that publications have on science and technology or in the general public, and guides funding decisions by governments and institutions.  Distilling the vast amount of available information and assigning meaning to such imprecise notions as ``publication impact'' is often accomplished by defining indicators.  An indicator is ``a statistical proxy of one or more metrics that allow for an assessment of a condition'' (National Research Council, 2012, p.11).\footnote{Definition from the National Center for Science and Engineering Statistics.}  Traditional bibliometric and scientometric indicators summarize the number of citations, publications, grants, or patents. More recently, the availability of social media data (e.g., Twitter, Wikipedia, Facebook, ResearchGate, Mendeley, etc.) has spurred the study of alternative metrics of publication impact, the so-called altmetrics (Haustein et al., 2015).

Common bibliometric and altmetric indicators are often based on notions of averaging
%over stratified data
and adding improvisational adjustments, but the field as a whole has not yet reached the statistical maturity of biostatistics.  Bornmann and Haunschild (2018a) drew on methodology used in biostatistics in motivating the altmetrics indicator MHq, which is based on Mantel-Haenszel estimators. The goal of these estimators is to describe the overall association between two binary variables based on stratified data. For example, Mantel-Haenszel estimators are used to study the association between two treatments (e.g., drug vs. placebo) and their outcomes (e.g., success or failure) when subjects are stratified by possible confounding factors (e.g., age groups or smoking status), or when combining the results of multiple studies in a meta-analysis. As a model for an indicator, the Mantel-Haenszel estimators have the advantage that the resulting statistics have well understood interpretations and inferential properties. The indicator MHq is relatively new, but it and its interpretation of data have already been extensively cited (Bornmann \& Haunschild, 2018b, 2018c, 2019; Bornmann, Haunschild, \& Adams, 2018, 2019; Copiello, 2020a, 2020b; Haunschild et al. 2019; Jiangbo et al., 2020;  Kassab, 2019;  Kassab et al., 2020;  Ortega, 2020; Tahamtan \& Bornmann, 2020; Wang, Lv, \& Hamerly, 2019).

This article corrects the interpretation and statistical inference for the MHq indicator and proposes a new indicator, MHRR.  Due to various reasons we explore in this article, Bornmann and Haunschild's interpretation of MHq is incorrect as they confuse the groups being compared and did not realize that MHq measures a risk ratio.
Furthermore, the variance estimator as used in the confidence interval proposed by Bornmann and Haunschild (2018a) is incorrect, and we provide a correct version.
%%%%While Bornmann and Haunschild's article is non-mathematical and does not explicitly derive any properties, both their interpretation and their confidence intervals contradict mathematical analysis.

There was already a brief exchange on the topic of the MHq indicator (Smolinsky \& Marx, 2018; Bornmann, Haunschild, \& Mutz, 2018; Smolinsky, 2019; Bornmann, Haunschild, \& Mutz, 2019).  In that exchange, we point out that the indicator is a type of risk ratio and raise the issue of the correctness of the variance formula.  Those who have followed this discussion can now immediately see that the Bornmann and Haunschild variance formula is incorrect in Section (\ref{sec:BHci}).  We also provide simulation results to evaluate the actual performance of the two variance estimators and the resulting confidence intervals under sparse data settings.  Furthermore in Section (\ref{sec:BHinterpretation}), we correct the interpretation that MHq compares mentioned articles of ``a unit" to the mentioned articles in the entire ``corresponding fields and publication years," as stated in the abstract of Bornmann and Haunschild (2018a).

We present the Mantel-Haenszel estimator of the row risk ratio (MHRR) as an alternative indicator to MHq, one that captures the phenomena MHq was expected to measure but does not. Interpretation and statistical inference for MHRR and other Mantel-Haenszel type estimators are well-established. Throughout, we use the small-world example given in Table 4 of Bornmann and Haunschild (2018a) to illustrate our points.

\section{\textbf{Mantel-Haenszel estimators for measuring association in stratified tables}}

\subsection{Stratified $2 \times 2$ contingency tables\nopunct} \label{sec:Tables} \hfill\quad

Consider data separated into $K$ strata numbered $i=1, \ldots, K$, where the data observed in stratum $i$ can be summarized in a $2\times 2$ table; see Table \ref{tab:cont}. For example, the data can consist of publication information of scientific articles (or other source items), stratified by subject area and/or publication year. Let group G refer to a specific set of articles, e.g., the ones published by a specific institution. The two rows in Table \ref{tab:cont} refer to whether an article belongs to group G (e.g., was published by a specific institution) or not, and the two columns refer to whether an article was mentioned (e.g., on Twitter) or not. The total number of articles included in stratum $i$ is $n_i = a_i + b_i + c_i + d_i$.  The purpose of the stratification is to control for confounding factors (e.g., publication year or subject area, as mentioned by Bornmann and Haunschild) that might have an influence on the frequency of being mentioned.

 \begin{table}[h!]
 \let\centering\relax
   \begin{flushleft}
      \captionof{table}{\\ \textbf{\textit {Contingency table}}} \label{tab:cont}
 \begin{tabular}{ r | c  c }
    \hline
    & \text{Mentioned} & \text{Not Mentioned} \\ \hline
    In G &$a_i$ & $b_i$ \\
    Not in G & $c_i$ & $d_i$\\
    \hline
  \end{tabular}\\
\textit{Note.}  Data from $i$-th stratum organized in a $2\times 2$ contingency table.\\
 \end{flushleft}
\end{table}

The organization of data for statistical analysis with Mantel-Haenszel estimators uses tables of the form of Table \ref{tab:cont}, which are $2\times 2$ contingency tables. They consist of the cross-classification of two binary categorical variables, e.g., whether an article belongs to group G or not (the row variable) and whether an article was mentioned or not (the column variable). The column sum $a_i + c_i$ then refers to the total number of mentioned articles in the $i$th stratum, while the column sum $b_i + d_i$ refers to the total number of articles not mentioned.

\subsection{Risk ratios, odds ratios, and Mantel-Haenszel statistics\nopunct} \label{sec:rrormh}
\hfill \quad

Two fundamental measures for describing the association between two binary categorical variables are the risk ratio (or relative risk) and the odds ratio (Agresti, 2019; Lachin, 2009). The risk ratio is the measure that is more familiar and easier to interpret and communicate. For articles in the $i$th stratum, the row risk ratio
\begin{equation}
\label{eq:tabRRR}
\text{RR}^r_i =  \frac {a_i/(a_i + b_i)} {c_i/(c_i + d_i)} % = \frac {a_i (c_i+ d_i)}{c_i(a_i + b_i)}
\end{equation}
compares the row proportions and describes how much more likely it is that an article is mentioned when it belongs to group G compared to when it does not belong to group G. Alternatively, comparing the column proportions in the $i$th stratum, the column risk ratio
\begin{equation}
\label{eq:tabCRR}
\text{RR}^c_i =  \frac {a_i/(a_i + c_i)} {b_i/(b_i + d_i)} % =  \frac {a_i (b_i+ d_i)}{ b_i(a_i + c_i)}
\end{equation}
describes how much more likely it is that an article belongs to group G when it is mentioned compared to when it is not mentioned. One usually only discusses the row risk ratio (and calls it the risk ratio), since the column risk ratio may be obtained as the row risk ratio from the transpose of Table \ref{tab:cont}. However, for our purposes, we simultaneously need both.

The odds of an article being mentioned in stratum $i$ are given by $a_i/b_i$ for articles belonging to group G, and by $c_i/d_i$ for articles not belonging to group G. The ratio of these two odds, the odds-ratio, is
\begin{equation}
\label{eq:tabOR}
\text{OR}_i =  \frac {a_i/b_i} {c_i/d_i} =  \frac {a_i/c_i} {b_i/d_i}.  %= \frac {a_i d_i}{b_i c_i}.
\end{equation}
The last expression in (\ref {eq:tabOR}) shows that we get the same result when comparing the odds of belonging to group G for articles mentioned (i.e., $a_i/c_i$) and not mentioned (i.e., $b_i/d_i$), so that a distinction between row and column odds ratios is not necessary.

The Mantel-Haenszel estimators combine the stratum specific risk ratios $RR_i$ (or the stratum specific odds ratios $OR_i$) into a single summary statistic. When the risk ratios (or odds ratios) across the strata are about the same, the Mantel-Haenszel estimator provides an estimate of the common association over the entire population. In particular, the Mantel-Haenszel estimators are weighted averages of the stratum-specific statistics, namely
\begin{equation}
\label{eq:generalweightedsums}
\frac{1}{w}\sum_{i=1}^K w_i \text{RR}_i \; \;  \text{ and } \; \;  \frac{1}{w}\sum_{i=1}^K w_i \text{OR}_i ,
\end{equation}
for the (row or column) risk ratio and odds ratio, respectively, where $w = \sum_{i=1}^K w_i$, and $w_i$ is a stratum specific weight.
 Each $\frac {w_i}w$ is a normalized weight and $\sum_{i=1}^K \frac {w_i}w =1$.

Specifically, the Mantel-Haenszel row risk ratio (\ref{eq:MHRRR}), column risk ratio (\ref {eq:MHCRR}) and odds ratio (\ref{eq:MHOR}) are computed as follows: % (Section 3.3 in Agresti \& Hartzel 2000).

\begin{eqnarray}
\label{eq:MHRRR}
\text{MHRR} &=&
\frac{1}{\sum_{i=1}^K \frac{c_i\left(a_i+b_i\right)}{n_i}} \sum_{i=1}^K \left( \frac{c_i\left(a_i+b_i\right)}{n_i}\right) \frac{a_i/(a_i+b_i)}{c_i/(c_i+d_i)}
=\frac{\sum_{i=1}^K \frac{a_i\left(c_i+d_i\right)}{n_i}}{\sum_{i=1}^K \frac{c_i\left(a_i+b_i\right)}{n_i}},\\
\label{eq:MHCRR}
\text{MHCR} &=&
\frac{1}{\sum_{i=1}^K \frac{b_i\left(a_i+c_i\right)}{n_i}} \sum_{i=1}^K \left( \frac{b_i\left(a_i+c_i\right)}{n_i}\right) \frac{a_i/(a_i+c_i)}{b_i/(b_i+d_i)}
=\frac{\sum_{i=1}^K \frac{a_i\left(b_i+d_i\right)}{n_i}}{\sum_{i=1}^K \frac{b_i\left(a_i+c_i\right)}{n_i}},\\
\label{eq:MHOR}
\text{MHOR} &=&
\frac{1}{\sum_{i=1}^K \frac{b_i c_i}{n_i}} \sum_{i=1}^K  \left(\frac{b_i c_i}{n_i}\right) \frac{a_i/b_i}{c_i/d_i}
=\frac{ \sum_{i=1}^K \frac{a_i d_i}{n_i}} {\sum_{i=1}^K \frac{b_i c_i}{n_i}}.
%\end{equation}
\end{eqnarray}

\noindent The middle term in each equation explicitly shows the weight $w_i$ that is used when forming the weighted average.  %Notice that the weights have random elements.  For example, in the row risk ratio (which is the usual Mantel-Haenszel risk ratio in the literature), the weight $w_i = \frac{C_i(A_i+B_i)}{N_i}$ has $A_i+B_i=a_i+b_i$ and $N_i=n_i$ fixed but $C_i$ is the value of a random variable.
The far right formula is the definition of each indicator.  Because these far right formulas involve ratio of sums (rather than a sum of individual ratios), they are applicable even for cases when the number of mentioned articles in some strata are zero, which is typical for sparse data. We note that Bornmann and Haunschild (2018a) denote the Mantel-Haenszel odds ratio MHOR by MHq'.

\section{\textbf{Interpretation and inference for Mantel-Haenszel type indicators}}
\label{sec:indicatorMHq}
\hfill \quad

\subsection{The indicator MHq\nopunct}
\label{sec:defMHq}
\hfill\quad

The indicator MHq (``Mantel-Haenszel quotient") as defined by Bornmann and Haunschild (2018a) can be written as
\begin{eqnarray}\label{eq:MHq}
\text{MHq} &=&
\frac{1}{\sum_{i=1}^K\frac{b_i\left(a_i+c_i\right)}{a_i + b_i + n_i}}\sum_{i=1}^K  \left(
\frac{b_i(a_i+c_i)}{a_i + b_i + n_i }\right) \frac{a_i/(a_i+c_i)}{b_i/(b_i+d_i)}
=\frac{\sum_{i=1}^K \frac{a_i\left(b_i+d_i\right)}{a_i + b_i + n_i }}{\sum_{i=1}^K\frac{b_i\left(a_i+c_i\right)}{a_i + b_i + n_i }}.
\end{eqnarray}
For the definition of MHq, Bornmann and Haunschild use the far right expression in Eq. (\ref{eq:MHq}), which equals the fraction $\frac{R}{S}$ in their Eq. (16).

We see from the middle expression of (\ref{eq:MHq}) that MHq is a weighted average of stratum-specific column risk ratios $RR^c_i$ just as MHRR and MHCR are weighted averages in  (\ref{eq:MHRRR}) and (\ref{eq:MHCRR}).  The Mantel-Haenszel estimator MHRR estimates the overall row risk ratio and the Mantel-Haenszel estimator MHCR estimates the overall column risk ratio when the values for individual strata are similar (Lachin, 2009).  Similarly,  we have that MHq estimates the overall column risk ratio.   In particular, as the sample size of each of the $K$ tables goes to infinity, MHq converges in probability to the population column risk ratio again when the individual strata have  similar risk ratios (homogeneity assumption).  We provide details in Theorem \ref{thm:convergence} in Appendix \ref{app:convergence}.

Note that the homogeneity assumption is not needed to define the indicators MHRR, MHCR, and MHq. They are a weighted sum of strata risk ratios, and the populations are always bounded.  Nevertheless, the indices only converge as theoretical populations increase to infinity under certain types of limits. (Davis, 1985).

Note that for the simplest case of just a single stratum (i.e., $K=1$), MHq reduces to the column risk ratio $\text{RR}^c_1$:
\begin{equation}
\label{eq:MHq1table}
\text{MHq}
%= \frac{a_1(b_1+d_1)}{b_1(a_1+c_1)}.
=\frac{ \frac{a_1\left(b_1+d_1\right)}{a_1 + b_1 + n_1 }}{\frac{b_1\left(a_1+c_1\right)}{a_1 + b_1 + n_1 }} =
 \frac{a_1/(a_1+c_1)}{b_1/(b_1+d_1)} = \text{RR}^c_1,
\end{equation}
which is the ratio of the proportion of mentioned articles that belong to group G to the proportion of not mentioned articles that belong to group G.

\subsection{Interpretation of the indicators\nopunct}\label{sec:toy}
\hfill\quad

We illustrate interpretation of the indicators using the same small-world example that Bornmann and Haunschild (2018a) present in their Table 4 (p. 1003), which we reproduce here as Table \ref{tab:BHdata}.

\begin{table}[h!]
 \let\centering\relax
   \begin{flushleft}
      \captionof{table}{\\ \textbf{\textit {Small-world example.}}} \label{tab:BHdata}
  \begin{tabular}{r  r | c  c }
\hline
  Publication set A &  &Articles
mentioned & Articles not
mentioned  \\
  \hline
 & Category 1 & 18 & 13 \\
 & Category 2 & 15 & 9  \\
 & Category 3 & 13 & 9   \\
 & Category 4 & 0 & 10  \\
 % \hline
Publication set B &  & &  \\
  \hline
 & Category 1 & 26 & 7 \\
 & Category 2 & 15 & 7  \\
 & Category 3  & 3 & 3   \\
 & Category 4  & 0  & 10  \\
  \hline
  \end{tabular}\\
\textit{Note.}  Data from Bornmann and Haunschild (2018a)\\
\end{flushleft}
\end{table}

We identify Publication Set B as group G (and hence Publication set A as not G), and the four Categories within a publication set as the $K=4$ strata. Using the notation of Table \ref{tab:cont}, Table \ref{tab:indicators} shows the counts $a_i, b_i, c_i$ and $d_i$ in each stratum and the values of the indicators defined in Eqs. (\ref{eq:MHRRR})--(\ref{eq:MHq}).

In Table \ref{tab:indicators}, notice that the 4-th stratum had no articles that were mentioned, i.e., $a_4 = c_4 = 0$ for this stratum.  Such strata are typically ignored when computing the statistic MHq or any of the Mantel-Haenszel estimators (\ref{eq:MHRRR})-(\ref{eq:MHq}), and it is omitted both here and in Bornmann and Haunschild (2018a).

\begin{table}[h!]
 \let\centering\relax
   \begin{flushleft}
      \captionof{table}{\\ \textbf{\textit {Point estimates and confidence intervals for indicators.}}} \label{tab:indicators}
 \begin{tabular}{r | r r r r r r r r}
\hline
  Stratum & $a_i$ & $b_i$ & $c_i$ & $d_i$ & $n_i$ & $\text{RR}^r_i$ & $\text{RR}^c_i$ & $\text{OR}_i$\\
  \hline
   $i=1$ & 26 & 7 & 18 & 13 & 64 & 1.36 & 1.69 & 2.68\\
   $i=2$ & 15 & 7 & 15 & 9  & 46 & 1.09 & 1.14 & 1.29\\
   $i=3$ & 3  & 3 & 13 & 9  & 28 & 0.85 & 0.75 & 0.69\\
   $i=4$ & 0  & 10 & 0 & 10  & 20 & \text{undefined} & \text{undefined}  & \text{undefined} \\
  \hline
  Indicator: & & & & & & MHRR=1.18   & MHCR=1.32 & MHOR=1.63 \\
             & & & & & &             & MHq=1.30 \\
  95\% C.I.: & & & & & & [0.91, 1.53]&  [0.85, 2.04] & [0.78, 3.39]\\
             & & & & & &             & [0.84, 2.00]$^*$ \\
  \hline
  \end{tabular}\\
\textit{Note.}  Each indicator is a weighted sum of the column above it.\\
 $^*$Computed using Eq. \ref{eq:SKM-CI}.
  \end{flushleft}
\end{table}

We now explicitly interpret the meaning of each of the indicators.  The intermediate probability step between the mathematical formula to the verbal description is given in Appendix \ref{app:interpretation}.
The indicator MHRR estimates the ratio of the chances of being mentioned for articles belonging to publication set B versus articles \textit {not belonging} to publication set B.\footnote{\textit {Not belonging to publication set B} is the same as \textit {belonging to publication set A} in this small-world example.  We use the phrase \textit {not belonging to publication set B} since it is more generally correct with more than two publication sets.}  Hence MHRR=1.18 indicates that, over all 4 categories, the chances of being mentioned are 18\% higher for articles that belong to publication set B compared to articles that do not belong to publication set B. In other words, articles from publication set B are 18\% more likely to be mentioned than articles not from publication set B.

The indicator MHCR estimates the ratio of the chances of belonging to publication set B for articles that are mentioned versus articles that are not mentioned. The value MHCR=1.32 indicates that, over all 4 categories, the chances of belonging to publication set B are 32\% higher for articles that are mentioned compared to articles that are not mentioned. In other words, mentioned articles are 32\% more likely to belong to publication set B.

As pointed out in Section \ref{sec:defMHq}, the indicator MHq is also an estimator of the overall column risk ratio. The value of MHq=1.30 indicates that the chances of belonging to publication set B are 30\% higher for articles that are mentioned compared to articles that are not mentioned. The MHq and the MHCR indicators differ only in the weights $w_i$ that are assigned to each stratum.  While MHq uses standardized weights of $w_1/w=0.414, w_2/w=0.402$ and $w_3/w=0.184$ for the three strata, MHCR uses weights $w_1/w=0.434, w_2/w=0.411$ and $w_3/w=0.154$ and assigns a relatively larger weight to the stratum with the largest sample size (stratum 1) and a relatively smaller weight to the stratum with the smallest sample size (stratum 3).

The indicator MHOR estimates the ratio of the odds of being mentioned between articles belonging to publication set B and those that do not. The value MHOR=1.63 indicates that the odds of being mentioned are 63\% higher for articles in publication set B.

\subsection{Statistical inference for MH\textnormal{\textbf{q}}\nopunct}
\label{sec:inference}
\hfill \quad

We now provide a formula for the estimate of the variance of $\ln(\text{MHq})$, which is used in the construction of its confidence interval. The derivations use similar assumptions and techniques as the computation of the estimate of the variance of the logarithm of the Mantel-Haenszel type estimators (Greenland \& Robins, 1985), but is complicated  by the fact that the stratum weights are not a linear function of the table entries. Details of the derivation are provided in Appendix \ref{app:Var[Estimators]}. We denote our estimator as $\text{Var}_\text{SKM}[\ln(\text{MHq})]$ and the estimator provided by Bornmann and Haunschild (2018a, Eq. (17)) as $\text{Var}_\text{BH}[\ln(\text{MHq})].$

Let
$$R_i =\frac{a_i (b_i+d_i)}{a_i+b_i+n_i},$$

$$S_i =\frac{b_i( a_i+c_i)}{a_i+b_i+n_i},$$

$$P_i=\frac{a_i(b_i+d_i)}{b_i(a_i+c_i)},$$

$$V_i=
(b_i+d_i)^2 \left( \frac {(n+b_i)^2}{(a_i+b_i+n_i)^4} \frac{a_ic_i }{a_i+c_i} +  \frac {a_i^2}{(a_i+b_i+n_i)^4}\frac{b_id_i }{b_i+d_i}\right),$$

$$W_i=
(a_i+c_i)^2 \left(  \frac {b_i^2}{(a_i+b_i+n_i)^4} \frac{a_ic_i }{a_i+c_i} +  \frac {(n_i+a_i)^2}{(a_i+b_i+n_i)^4}\frac{b_id_i }{b_i+d_i}
\right),$$

$$Q_i=  - \frac{(a_i+c_i) (b_i+d_i)}{(a_i+b_i+n_i)^4} \left(\frac{a_i b_i c_i (b_i+n_i)}{a_i+c_i}+\frac{a_i b_i d_i (a_i+n_i)}{b_i+d_i}\right),$$

and let $R=\sum_{i=1}^K  R_i$ and $S=\sum_{i=1}^K  S_i$.

Our estimator for the variance of $\ln(\text{MHq})$ is
\begin{equation}
\label{varestimator2}
\text{Var}_\text{SKM}[\ln(\text{MHq})]
 = \frac1{RS} \sum_{i=1}^K  ( \frac 1{P_i} (V_i  + R_i^2 ) - 2 (Q_i + R_iS_i) + P_i(W_i + S_i^2 )    ).
\end{equation}

Our variance formula allows for the computation of confidence intervals for MHq.  The 95\% confidence interval has endpoints

\begin{equation}
\label{eq:SKM-CI}
\exp \left(\ln(\text{MHq}) \pm 1.96 \sqrt{\text{Var}_\text{SKM}[\ln (\text{MHq})] } \right).
\end{equation}

Applied to the data from Table \ref{tab:indicators}, we find a 95\% confidence interval for the MHq indicator of $[0.84, 2.00]$. With 95\% confidence, the chances of belonging to publication set B are between 16\% lower to 100\% higher for articles that are mentioned compared to articles that are not mentioned.

\subsection{Statistical inference for MHRR, MHCR and MHOR\nopunct}
\label{sec:inferencepsi}
\hfill \quad

Variance estimators for the Mantel-Haenszel indicators are well-established (Agresti \& Hartzel, 2000). The variance estimator for the logarithm of MHRR is given by
$$
\mbox{Var}[\ln(\text{MHRR})] = \frac{ \sum_{i=1}^K{((a_i+b_i)(c_i+d_i)(a_i+c_i) - a_ic_in_i)/n_i^2}}{(\sum_iR_i ) (\sum_i S_i)},
$$
where now $R_i = a_i(c_i+d_i)/n_i$ and $S_i=c_i(a_i+b_i)/n_i$. This leads to the 95\% confidence interval of the form
$$
\exp \left(\ln(\text{MHRR}) \pm 1.96 \sqrt{\mbox{Var}[\ln(\text{MHRR})]} \right).
$$

The variance estimator for the logarithm of the Mantel-Haenszel column risk ratio uses the same formulas, but applied to the transposed $2 \times 2$ tables, i.e., replacing $b_i$ with $c_i$ and $c_i$ with $b_i$ in all expressions above.

%Since the standard Mantel-Haenszel relative risk is row relative risk, the variance estimator for the logarithm of the Mantel-Haenszel column relative risk is given by switching the rows and columns:
%$$
%\widehat{\mbox{Var}}[\ln(\hat{\psi}_{MH}^C)] =\frac{ \sum_{i=1}^K{((a_i+c_i)(b_i+d_i)(a_i+b_i) - a_ib_in_i)/n_i^2}}{( \sum_iR_i ) ( \sum_i S_i )},
%$$
%where $R_i = a_i(b_i+d_i)/n_i$ and $S_i=b_i(a_i+c_i)/n_i$. This leads to the 95\% confidence interval of the form
%$$
%\exp \left(\ln(\hat{\psi}_{MH}^C) \pm 1.96 \sqrt{\widehat{\mbox{Var}}[\ln(\hat{\psi}_{MH}^C)]} \right).
%$$

For the small-world data in Table \ref{tab:indicators}, MHRR=1.18 and the 95\% confidence interval equals [0.91, 1.53]. With 95\% confidence, we estimate that, overall, articles from publication set B are between 9\% less and 53\% more likely to be mentioned, compared to articles not belonging to publication set B. The value of MHCR=1.32, with 95\% confidence interval equal to [0.85, 2.04]. With 95\% confidence, we estimate that, overall, mentioned articles are between 15\% less and 104\% more likely to belong to publication set B, compared to articles not mentioned.

For completeness, the variance estimator for the logarithm of MHOR is
\begin{eqnarray*}
\mbox{Var}[\ln(\text{MHOR})] &=& \frac{ \sum_{i=1}^K (a_i+d_i)(a_id_i)/n_i^2} {2\left(\sum_{i=1}^K a_id_i/n_i\right)^2} +
\frac{ \sum_{i=1}^K (b_i+c_i)(b_ic_i)/n_i^2} {2\left(\sum_{i=1}^K b_ic_i/n_i\right)^2}\\
& & + \frac{ \sum_{i=1}^K [(a_i+d_i)(b_ic_i) + (b_i+c_i)(a_id_i)]/n_i^2} {2\left(\sum_{i=1}^K a_id_i/n_i\right)\left(\sum_{i=1}^K b_ic_i/n_i\right)},
\end{eqnarray*}
resulting in a 95\% confidence interval for MHOR=1.63 of [0.78, 3.39]. With 95\% confidence, we estimate that, over all 4 categories, the odds of being mentioned are between 22\% lower and 239\% higher for articles belonging to publication set B compared to articles that do not.

\section{\textbf{Corrections to Bornmann and Haunschild}}
\label{sec:BH}

Bornmann and Haunschild (2018a) do not organize the stratified publication data in the form of $K$ $2\times2$ contingency tables in the form of Table \ref{tab:cont}, but use what they call \emph{cross-tables} of the form of Table \ref{tab:notcont}; see Table 3 in Bornmann and Haunschild (p. 1002). While the first row of Table \ref{tab:notcont} also refers to the frequencies of mentioned and not mentioned articles belonging to publication set B, the second row refers to the total number of mentioned and not mentioned articles in the \emph{world}. The world refers to all articles, i.e., those belonging to group G and those not belonging to group G. These are the column sums: $a_i+c_i$ for the mentioned and $b_i+d_i$ for the not mentioned articles in Table \ref{tab:cont}. Bornmann and Haunschild write that the motivation to use frequencies for articles in the world in the second row is that in bibliometrics it is conventional to use the world (i.e., all articles published) as a reference set when making comparisons.

\begin{table}[h!]
 \let\centering\relax
   \begin{flushleft}
      \captionof{table}{\\ \textbf{\textit {Cross-table}}} \label{tab:notcont}
   \begin{tabular}{  r | c  c  }
    \hline
    & \text{Mentioned} & \text{Not Mentioned} \\ \hline
    In G &$a_i$ & $b_i$\\
   In World & $a_i+c_i$ & $b_i+d_i$\\
    \hline
  \end{tabular}\\
\textit{Note.}  Data from $i$-th stratum organized in a cross-table.\\
 \end{flushleft}
\end{table}

\subsection{Bornmann and Haunschild's interpretation\nopunct}
\label{sec:BHinterpretation}\hfill \quad

We showed in Section \ref{sec:defMHq} that MHq is an estimator of the overall column risk ratio and needs to be interpreted as such. This is why an MHq of 1.30 indicates that mentioned articles are 30\% more likely to belong to publication set B than not mentioned articles.  In their verbal interpretations of MHq, Bornmann and Haunschild (2018, Section 2.3) misinterpret MHq in terms of a row risk ratio that compares publication set B to the world. For instance, Bornmann and Haunschild write that MHq=1.30 indicates ``\ldots the publications in set B have 30\% higher chances for being mentioned than the world’s publications." (2018, p. 1003). This interpretation would imply that MHq is a ratio of two row percentages, which we have shown is not the case. Rather, MHq compares column percentages, and hence compares mentioned articles with not mentioned articles in terms of the proportion belonging to publication set B.

\subsection{Bornmann and Haunschild's confidence interval\nopunct}
\label{sec:BHci}\hfill \quad

Bornmann and Haunschild's variance estimator $\text{Var}_\text{BH}[\ln (\text{MHq})]$ used in the construction of their confidence interval for MHq is given in their Eq. (17). The easiest way to see that their formula is problematic is to consider the special case of just a single stratum, i.e., $K=1$, where MHq equals the simple column risk ratio $RR^c$. It is well-known (Katz et al., 1978) that the estimated variance of the logarithm of the risk ratio is given by
\begin{equation}
\label{eq:varRR}
\text{Var}[\ln(\text{MHq})] = \frac{1}{a_1} -\frac{1}{a_1+c_1}+\frac{1}{b_1}-\frac{1}{b_1+d_1}.
\end{equation}
It is straightforward to check that our variance formula $\text{Var}_\text{SKM}[\ln(\text{MHq})] $ reduces to this expression for the special case of $K=1$. However, Bornmann and Haunschild's variance formula yields, for $K=1$,
\begin{equation}
\label{eq:1tablevar}
\text{Var}_\text{BH}[\ln(\text{MHq})] = \frac{1}{a_1} +\frac{1}{a_1+c_1}+\frac{1}{b_1}+\frac{1}{b_1+d_1},
\end{equation}
clearly overestimating the variance of the log-risk ratio.

To compare Bornmann and Haunschild's variance formula to our variance formula for larger values of $K$, we ran a simulation study, which shows that Bornmann and Haunschild's formula considerably overestimates the true variance and leads to confidence intervals that are too wide. On the other hand, our variance formula results in confidence intervals for MHq that have coverage probability very close to the nominal level, and which are considerably narrower, allowing for more precise inference.

\section{\textbf{Simulation study}}\label{sec:simresults}\hfill \quad

\begin{figure}[h]
 \let\centering\relax
   \begin{flushleft}
 \caption{\\ \textbf{\textit {Bias in the estimation of the standard deviation}}}
     \label{fig:BiasVar1}
  \includegraphics[width=400pt]{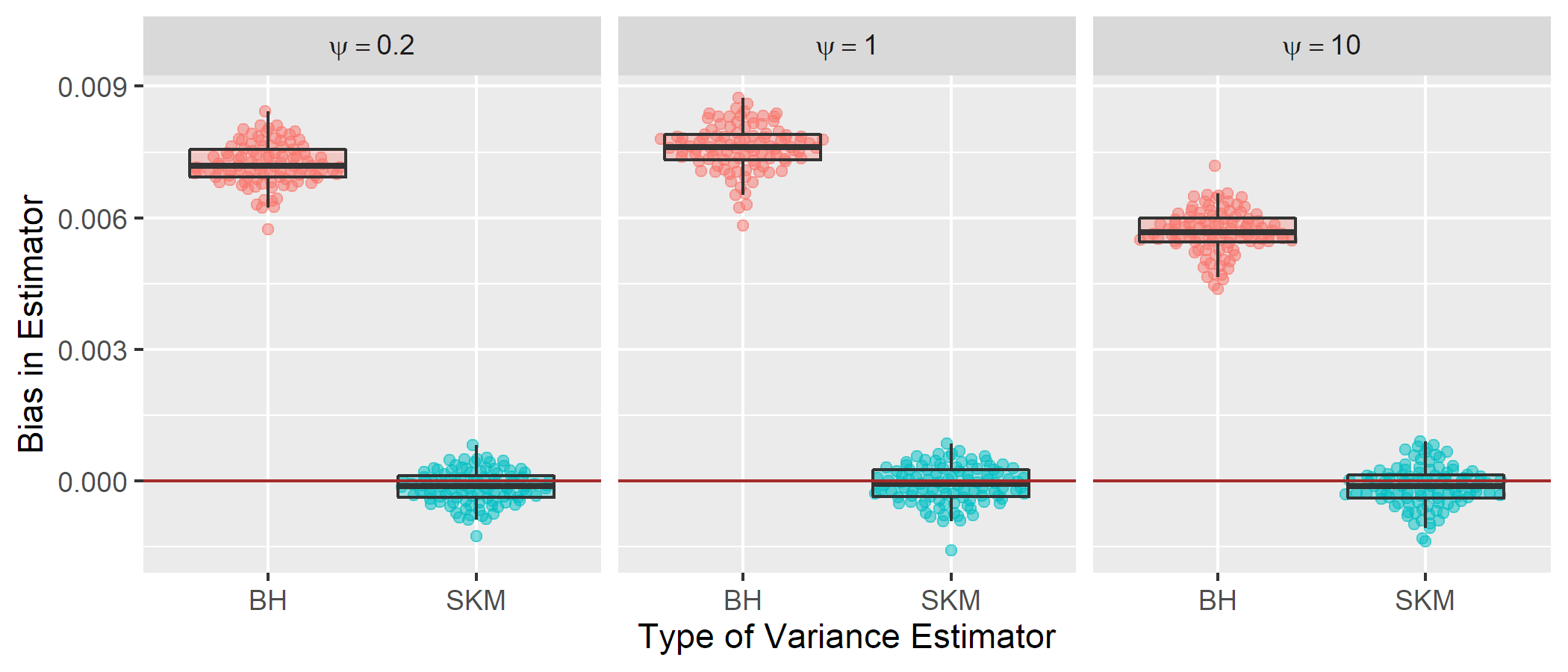}\\
\textit{Note.} Bias in the estimation of the standard deviation of ln(MHq) using $\text{Var}_\text{SKM}[\ln(\text{MHq})]$ and $\text{Var}_\text{SKM}[\ln(\text{MHq})]$ with K= 30.
 \end{flushleft}
\end{figure}

To assess Bornmann and Haunschild's (BH) variance formula and our (SKM) variance formula for MHq in practice, we conducted a simulation study. We simulated data sets (under a column binomial model) of $K=30$ strata, each with a fixed sample size of 100 articles in the mentioned group and 1000 articles in the not mentioned group. We kept these sample sizes the same across all 30 strata. Of the 100 articles in the mentioned group, we assumed a proportion of $p_{1i}$ are from group G, where we let $p_{1i}$ vary according to a uniform distribution between 1\% and 20\% across the $i=1,\ldots,30$ strata. Of the 1000 articles in the not mentioned group, we assumed a proportion of $p_{2i}$ were from group G, where $p_{2i}=p_{1i}/\psi$. This ensured that the strength of the association $\psi = p_{1i}/p_{2i}$ was the same in each stratum, as assumed by the Mantel-Haenszel method. We looked at three possible values for $\psi$: $\psi=0.2$ (overall, mentioned articles were 80\% less likely to be from group G compared to not mentioned articles), $\psi=1$ (no association; overall, mentioned and not mentioned articles were equally likely to be from group G), and $\psi=10$ (overall, mentioned articles were 10 times more likely to be from group G compared to not mentioned articles). With these underlying population parameters fixed (which gave rise to sparse data), we generated 10,000 datasets, computing $\ln(\text{MHq})$ from each. This provided 10,000 realizations of $\ln(\text{MHq})$ under the given parameter settings, and the variance of those 10,000 values provided the ground truth for the variability in measuring $\ln(\text{MHq})$. This ground truth was compared to the BH and SKM variance formulas, using the true parameters instead of the estimated ones when evaluating those formulas. The difference between the values obtained via the formulas and the ground truth was an estimate of the bias in the variance formula. We summarize the results of our simulations in Figure \ref{fig:BiasVar1}, where a single dot in the plot represents one such difference. For better interpretability, we plot the difference in the standard deviations instead of the variances.

We repeated the entire procedure above (drew a new set of $p_{1i}$'s from the Uniform$(0.01,0.2)$ distribution and generated a new set of 10,000 datasets of 30 strata each) 99 more times and obtained a total of 100 such differences between the actual standard deviation and the one predicted by the formulas. If Bornmann and Haunschild's variance formula were unbiased, we would have expected those 100 differences to cluster around 0. However, as is evident from Figure \ref{fig:BiasVar1}, there was considerable (positive) bias in the BH variance formula, as was expected based on the case for $K=1$. On the other hand, the simulations indicated the SKM formula was nearly unbiased, with the 100 differences closely clustering around 0.

We conducted several other simulation studies, where we varied the underlying simulation parameters, such as investigating smaller and larger values of $K$ or the sample sizes of mentioned and not mentioned articles, and found similar results regarding the bias.

\begin{figure}[h]
 \let\centering\relax
   \begin{flushleft}
 \caption{\\ \textbf{\textit { Coverage rate of nominal 95\% confidence intervals}}}
     \label{fig:BiasVar2}
  \includegraphics[width=400pt]{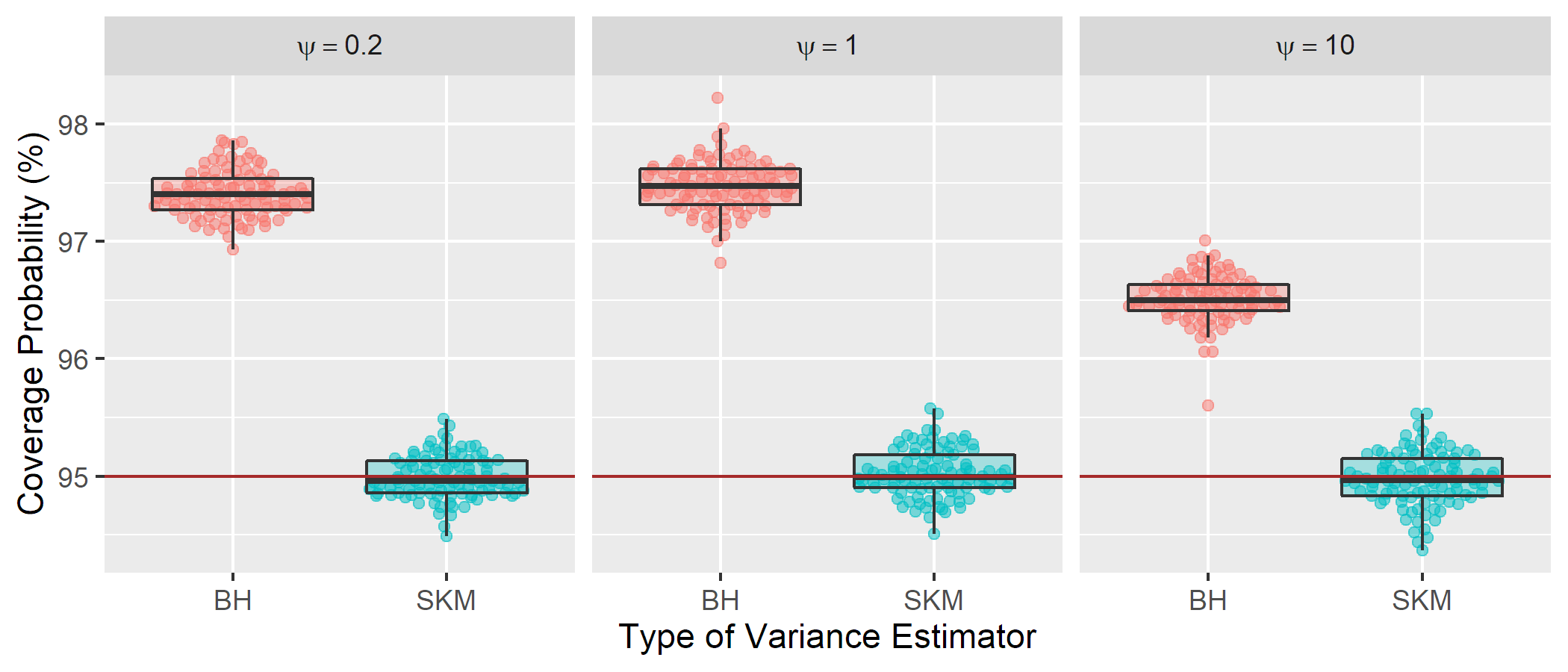}\\
\textit{Note.} Coverage rate of nominal 95\% confidence intervals for MHq using $\text{Var}_\text{BH}[\ln(\text{MHq})]$ and $\text{Var}_\text{SKM}[\ln(\text{MHq})]$ with K= 30.
 \end{flushleft}
\end{figure}

We also conducted a simulation study investigating the coverage rate of the 95\% confidence interval for MHq. Figure \ref{fig:BiasVar2} shows the coverage rate when $K=30$.  The sample size in the mentioned group was 100, and the sample size in the not mentioned group was 1000 in each stratum. Each dot in the plot corresponds to a particular setting of the underlying probabilities $p_{1i}$ of belonging to group G for articles mentioned and not mentioned, and a particular value of $\psi$. We again generated 10,000 datasets, computed the confidence interval for MHq from each simulated dataset (using the BH or SKM formula), and then checked if the interval contained the MHq value used to generate the data. In theory, for 9500 of the 10,000 data sets generated (95\%), the resulting confidence interval should have contained the true MHq. Each dot in Figure \ref{fig:BiasVar2} represents the actual coverage rate (the percent of the 10,000 intervals that actually covered the true MHq) under one particular setting of population parameters. We see that with the BH variance formula the actual coverage rate tends to be above the nominal 95\% level, indicating that the confidence intervals were too wide. For intervals using the SKM variance formula, the coverage rate tends to hover around the nominal 95\% level, as it should. We also created a plot (not shown) similar to Figure \ref{fig:BiasVar2}, plotting the average length of the 10,000 generated intervals over 100 different population parameter settings. The plot and simulation results indicate that the BH intervals are wider, on average, by 13\% when $\psi=0.2$, 12\% when $\psi=1$ and 7\% when $\psi=10$ compared to intervals using the SKM variance formula.

\section{\textbf{Discussion}}

In defining and discussing MHq, Bornmann and Haunschild (2018a) use the Mantel-Haenszel odds ratio formula and Robins, Breslow, and Greenland's variance formula for the log-odds ratio (Robins et al., 1986) taken from the book by Fleiss et al. (2003). However, those formulas, like all formulas concerning Mantel-Haenszel estimators appearing in the statistical literature, have variables specifically referring to contingency table entries and not to table entries in other cross-tables.  As a consequence of using cross-tables, all Mantel-Haenszel type indicators (such as MHq) lose their original meaning and interpretation. It is perhaps surprising that the incorrect application of the odds ratio formula results in a meaningful estimator for the column risk ratio.

The indicator MHq is a meaningful statistic to measure mentions to a group G, although the original interpretation was incorrect.  MHq compares the percent of mentioned articles that belong to group G to the percent of not mentioned articles that belong to group G.  It does not, however, give a direct comparison to the world.  Furthermore, researchers who use Bornmann and Haunschild's confidence interval for MHq have untrustworthy inference. As our simulations showed, those confidence intervals for MHq are too wide meaning that MHq is more precise than previously believed.

Thelwall's (2017a, 2017b) indicators MNPC and EMNPC are compared to MHq in a study of peer recommendations from F1000Prime on three groups (G= Q0, Q1, and Q2) by Bornmann and Haunschild (2018a).  Bornmann \& Haunschild give a comparison of the three indicators on the three groups in their Figure 1 (p. 1007).  Since they used the wrong confidence interval formula, the correct confidence intervals are likely narrower and would improve their result.  Bornmann and Haunschild observe that values for MNPC and EMNPC cluster very near 1 for all groups, but that is not true for MHq.  The fact that MHq does not include a direct comparison to the world gives it greater separation in value (see Appendix \ref{app:interpretation}, Eq \ref{eq:worldcompcolumn}).

The indicator MHRR captures the notion that MHq was originally intended to measure.   It is a transparent measure since it directly describes how much more (less) likely articles in group G are mentioned compared to articles not in group G.  Furthermore, if one also knows the fraction of the world's articles that belong to group G, then one may easily obtain how much more (less) likely articles in group G are mentioned compared to articles in the world (Appendix \ref{app:interpretation}, Eq. \ref{eq:worldcomprow}), i.e., the ratio of percentage of mentioned  articles in G to mentioned articles in the world.

The range of possible values for each of MHRR, MHq, and MHCR is $[0,\infty)$.  A value of 1 indicates consistent performance between groups G and G$^C$, and hence between group $G$ and the world.  Values greater (less) than 1 indicate better (worse) performance in group G than in the world. Different subgroups in the world may be compared directly by the values of the indicator for any of MHRR, MHq, and MHCR.

Bornmann and Haunschild (2018a) also discuss an alternative indicator called MHq' that is identical to the Mantel-Haenszel odds ratio MHOR. The odds ratio is primarily of interest because it approximates the risk ratio when probabilities are small (Greenland \&  Robins, 1985). This approximation is valid when G is a small part of the world and the probabilities for articles being mentioned are small regardless of belonging to G across strata. In this situation, MHRR, MHCR, MHq, and MHq' will be similar in value. The usefulness of this approximation is limited, and as Greenland and Robins assert, ``in typical cohort studies this approximation may no longer hold" (Greenland \& Robins, 1985, p. 55).   The indicators MHRR and MHq (or MHCR) compute the row and column risk ratio directly without the need for an approximation and are not prone to misinterpretation by confusing odds and probabilities.

The indicators MHRR and MHq are easy to interpret if the risk ratio does not vary dramatically from stratum to stratum, but are still valid and meaningful indicators under heterogeneity.  This analysis was done by Noma and Nagashima (2016), who showed that Mantel-Haenszel estimators still provide reasonable and intuitive summary statistics.

 \subsection{Comparisons to the world\nopunct}\label{furtherresearch} \hfill \quad

Although bibliometrics researchers often compare results to the world or normalize by world results, we believe that using risk ratios result in a better indicator.  Risk ratios and odds ratios are used by statisticians as the comparison groups are distinct, i.e., categorical.
Comparisons of a group G to the world may confound the size of the group with the probability of being mentioned. If the size of group G is small compared to the world, then the comparison to the world approximates the risk ratio.  However, as the size of group G gets larger, the comparison with the world simply approaches 1 and becomes insensitive as a measure of performance. We discuss this in Appendix \ref{app:interpretation} (Eq \ref{eq:worldcompcolumn}), which shows explicitly that the size of G is a confounding factor. We believe this is a good reason to use the risk ratio (e.g., our indicator MHRR, but also MHq) over a comparison to the world.

The situation for comparisons with the world becomes worse when the purpose of stratification is to avoid phenomena like Simpson’s paradox. It is worse because comparisons to the world now depend on the relative size of group G in each stratum. This is not the case for methods using Mantel-Haenszel statistics (e.g., all indicators mentioned in this article), because they are all relative measures, where the size of group G does not play a role.

\subsection{Limitations of the indicators\nopunct}\label{limitations} \hfill \quad

If there is substantial heterogeneity in the risk ratios across the strata, the actual coverage probabilities of confidence intervals that are computed based on the assumption of homogeneity suffer and may fall well below the nominal 95\% level, as indicated in several simulation studies (e.g., Klingenberg, 2014; Böhning, Sangnawakij and Holling, 2021). The latter reference discusses the bootstrap for finding a confidence interval for the Mantel Haenszel risk ratio.

A substantial limitation for MHq, MHRC, and any comparison to the world is the required sample size.  However, this is not an issue for MHRR.  One may use the entire world reference set as the sample (i.e., the entire population) and all indicators are meaningful, but one must be cautious using samples.

We illustrate the issue with the sample size.  Suppose one wishes to evaluate mentions of articles from University X compared to all university articles in the European Union, and University X is typical among the 2,725 universities in the EU listed in uniRank (Universities and Higher Education in Europe, 2020).  For MHRR, one may take a random sample of 100 articles from University X and a random sample of 100 articles from other EU universities.  This method uses binomial sampling in each row and is valid for studying the row risk ratio (Agresti, 2019).  It has the advantage that one may directly select articles from University X and use a small overall sample (here 200 articles).  For MHq or MHRC, this is not sufficient and there are two options both requiring large samples.  One option is to take a large random sample from EU university articles without bias toward selection from University X or mentioned articles.  This is an example of multinomial sampling (Agresti, 2019).  However, to ensure that the sample contains at least 100 articles from University X, one would need to sample on the order of $2,725\times 100= 272,500$ articles.   A second option for MHq or MHRC is a binomial model using binomial sampling in each column.  The columns refer to mentioned and not mentioned articles, so the selection must still be unbiased toward selection from University X.  A large random sample of mentioned and not mentioned articles is still required to accumulate 100 articles from University X by chance.  Similarly, for a direct comparison to the world, multinomial sampling is required.

%We are currently exploring such indicators that are in the spirit of Mantel-Haenszel type estimators.

 \subsection{Further research\nopunct}\label{furtherresearch} \hfill \quad
The indicator MHq is similar to MHCR.  We cannot say which of MHCR and MHq is superior.
Variance and convergence validity need to be studied.  In the one example we computed in Table \ref{tab:indicators}, MHq and MHCR had similar values and confidence intervals.

MHRR and MHq are distinct statistics.  We hope they can be further investigated, based on their properties discussed in this article, with regards to important properties for indicators in bibliometrics, such as convergence validity.

\section{\textbf{Acknowledgements}}

The authors are grateful to the reviewers for their careful reading and helpful suggestions to improve the article.

%%%%%%%%%%%%.  Appendix begins here %%%%%%%%

\vfill
\pagebreak
\appendix

\section{}\label{app:convergence}

\begin{center}
\textbf{Convergence of MHq}
\end{center}

In these appendices, we use more conventional mathematical notation.  Let $A_i$ and $B_i$ denote the random variables that are the entries of the $i$-th contingency table and let $a_i$ and $b_i$ denote their observed values. The random variables $A_i$ and $B_i$ are binomial with corresponding parameters $p_{1i}$ and $p_{2i}$ and sample sizes $n_{1i}$ and $n_{2i}$ for $i= 1, ..., K$. Therefore, $E[A_i] = p_{1i}n_{1i}$,  $E[B_i] = p_{2i}n_{2i}$, $\text{Var} [A_i] = p_{1i}(1-p_{1i})n_{1i}$, $\text{Var} [B_i] = p_{2i}(1-p_{2i})n_{2i}$ with corresponding estimators $\hat p_{1i}=\frac{a_i}{n_{1i}}$, $\hat p_{2i}=\frac{b_i}{n_{2i}}$, $\hat E[A_i]=a_i$ and $\hat E[B_i]=b_i$.  Let $\psi$ be the column risk ratio and $\psi_i$ the column risk ratio of the $i$-th table.  The homogeneity assumption common to all Mantel-Haenszel type estimators is that $\psi_i =\psi$ (Lachin, 2009). An estimate of $\psi$ is $\hat\psi_i$  for each $i$.  Items with different subscripts are from different tables and are independent.  Write $P(E)$ for the probability of an event $E$, and $E^C$ for the complement.

We consider multi-indexed sequences of random variables, $Y_{\vec n}$, where $\vec n = (n_{11}, n_{21},\cdots, n_{2K})$.  The $\hat{\psi}_i$ are such sequences of random variable, but we usually suppress the $n$ indices,  $\hat\psi_i = \hat\psi_{i, n_{1i}, n_{2i}}=\hat\psi_{i,\vec n}$.  Similarly, the normalized weights $\frac{w_i}{\sum_{j=1}^K w_j} = X_i$ are too:  $X_i = X_{i,\vec n}$.  They satisfy $\sum_{i=1}^K X_{i,\vec n} = 1$.  A sequence of random variables must be defined on a single sample space.  That is the case for the $\psi_i$'s and the $X_i $'s.\footnote{The sample space is the $2K$ fold Cartesian product $\mathbb S = \mathbb S_{p_{11}} \times \cdots \times \mathbb S_{p_{2K}}$ where $\mathbb S_{p_{ij}}$ is the sample space for the infinite coin flip of a coin with probability of a head ${p_{ij}}$, which is used for binomials with arbitrarily large samples (Williams, 1991, p.24, Ex 2.3b).}

Recall that a sequence of random variables $Y_{\vec n}$ converge in probability to $Y$ or $Y_{\vec n} \overset{p}{\to} Y$ means for every $\epsilon>0$, $\lim_{n_{11}, n_{21},\cdots, n_{2K} \to \infty } P( |Y_{\vec n} - Y|<\epsilon) = 1$.  Explicitly expressing the limit to infinity,  $Y_{\vec n} \overset{p}{\to} Y$ means:

$\forall \epsilon, \eta >0, \exists N$ such that
$n_{11}, n_{21},\cdots, n_{2K}> N \implies | P(|Y_{\vec n}-Y|<\epsilon) -1 | < \eta$
(Kudryavtsev, 1989).
 The last expression $ | P(|Y_{\vec n}-Y|<\epsilon) -1 | < \eta$ is the same as
 \begin{equation}
 \label{eq:epsilon-eta}
 P(|Y_{\vec n}-Y|<\epsilon) > 1- \eta \ \text{ or }\ P(|Y_{\vec n}-Y|\ge \epsilon) < \eta
 \end{equation}
We recall that $\hat\psi_i \overset{p}{\to} \psi $ as $n_{i,j}\to\infty$ and we have suppressed the subscripts indicating the dependence on the $n$'s, e.g., $\vec n$.  The following theorem establishes that MHq converges in probability to the common risk ratio.

 \begin{theorem}
Suppose $X_{i,\vec n}$ are a sequence of random variables for $i=1,\cdots, K$ and $\vec n = (n_{11}, n_{21},\cdots, n_{2K})$ with $n_{ji}\in\mathbb N$ and suppose for each $\vec n$, $\sum_{i=1}^K X_{i,\vec n} = 1.$  Let $\hat\psi_{i,\vec n} = \frac{A_i n_{2i}}{B_i n_{1i}}$ be the column risk estimator for the $i$-th table.  Then, $\sum_{i=1}^K X_{i,\vec n}\hat\psi_{i,\vec n} \overset{p}{\to}\psi$.
 \label{thm:convergence}
 \end{theorem}

 \begin{proof}. Suppose we are given $\epsilon, \eta>0$, which are now fixed.  We must produce an $N$ so that
 \begin{equation} \label{eq:requirement}
 n_{11}, n_{21},\cdots, n_{2K}>N \implies P(|\sum_{i=1}^K X_{i,\vec n}\hat\psi_{i,\vec n}-\psi |<\epsilon) > 1- \eta.
  \end{equation}
 Note that  $P(|\sum_{i=1}^K X_{i,\vec n}\hat\psi_{i,\vec n}-\psi |<\epsilon)|$ is shorthand for $P(B)$ where
  $B=\{ x |  P(|\sum_{i=1}^K X_{i,\vec n}(x)\hat\psi_{i,\vec n}(x)-\psi |<\epsilon) \}$.

Since $\hat\psi_{i,\vec n} \overset{p}{\to} \psi$, we know that there are $N_i$'s such that
 $ n_{11}, n_{21},\cdots, n_{2K}>N_i \implies P(|\hat\psi_{i,\vec n}-\psi |<\epsilon)|) > 1- \frac{\eta}{2K}$ or equivalently
 $P(|\hat\psi_{i,\vec n}-\psi |\ge \epsilon)|) < \frac{\eta}{2K}$.  We now choose $N=\text{max}\{ N_i \}$.

 Let $A_{i \vec n}= \{ x |  P(|\hat\psi_{i,\vec n}(x)-\psi |<\epsilon) \}$, so $A_{i \vec n}^C = \{ x |  P(|\hat\psi_{i,\vec n}(x)-\psi |\ge \epsilon) \}$, $P(A_{i \vec n}) = P(|\hat\psi_{i,\vec n}-\psi |<\epsilon)$, and $P(A_{i \vec n}^C) = P(|\hat\psi_{i,\vec n}-\psi |\ge \epsilon)$.

 For $n_{11}, n_{21},\cdots, n_{2K}>N$, we have
 \begin{equation}
 \begin{aligned}\label {eq:DeMorgan}
 P(\forall i \ |\hat\psi_{i,\vec n}-\psi |<\epsilon)) &= P(\cap A_{i \vec n})\\
 &= P((\cup A_{i \vec n}^C)^C),\quad \text {by De Morgan's law}\\
 &= 1- P(\cup A_{i \vec n}^C), \quad \text {the complement} \\
&\ge 1-  \sum_{i=1}^K P(A_{i \vec n}^C), \quad \text {by the principle of inclusion/exclusion}\\
 &\ge 1- \sum_{i=1}^K \frac{\eta}{2K} > \ 1-\eta
 \end{aligned}
 \end{equation}

 We next show that
  \begin{equation}
  \begin{aligned}\label{eq:subset}
  & \forall i \ |\hat\psi_{i,\vec n}-\psi |<\epsilon \implies P(|\sum_{i=1}^K X_{i,\vec n}\hat\psi_{i,\vec n}-\psi |<\epsilon)\\
   & \text{ or equivalently, } P(\cap A_{i \vec n}) \subset P(B).
    \end{aligned}
  \end{equation}
  This statement follows because
   \begin{align*}
  |\sum_{i=1}^K X_{i,\vec n}\hat\psi_{i,\vec n}-\psi | &= |\sum_{i=1}^K X_{i,\vec n}\hat\psi_{i,\vec n}-\sum_{i=1}^K X_{i,\vec n}\psi | \\
  & \le \left( \sum_{i=1}^K X_{i,\vec n} \right) |\hat\psi_{i,\vec n}(x)-\psi | \\
  & \le \left( \sum_{i=1}^K X_{i,\vec n} \right) \epsilon \quad =  \epsilon.
   \end{align*}

 We complete the proof using (\ref{eq:subset}) and (\ref{eq:DeMorgan}): For $n_{11}, n_{21},\cdots, n_{2K}>N,$
     \begin{align*}
 P(|\sum_{i=1}^K X_{i,\vec n}\hat\psi_{i,\vec n}-\psi |<\epsilon) &= P(B) > P(\cap A_{i \vec n}) > 1- \eta,
    \end{align*}
 which establishes (\ref{eq:requirement}).
 \end{proof}

\vfill
\pagebreak

\section{}\label{app:interpretation}

\begin{center}
\textbf{Interpretation}
\end{center}

Verbal interpretations for the various Mantel-Haenszel type indicators rely on specifying the appropriate conditional probabilities in the definition of the risk ratio (or odds ratio). For a randomly selected article, let G indicate the event of belonging to group G and let M indicate the event of being mentioned. Let $i$ denote that an article belongs to the $i$-th stratum. We use the superscript $C$ to denote the complement event. The probabilities below can be interpreted as the actual probabilities for population data, or their estimates based on a random sample from the population. Using notation from Table \ref{tab:cont}, let $P(G|i)= \frac{a_i+b_i}{n_i}$, $P(G^C | i)= \frac{c_i+d_i}{n_i}$, $P(M|i)= \frac{a_i+c_i}{n_i}$, $P(M^C|i)= \frac{b_i+d_i}{n_i}$, $P(G|M,i)= \frac{a_i}{a_i+c_i}$,  $P(G|M^C,i)= \frac{b_i}{b_i+d_i}$,  $P(M|G,i)= \frac{a_i}{a_i+b_i}$,  and $P(M|G^C,i)= \frac{c_i}{c_i+d_i}$.

The row risk ratio in stratum $i$ is
$$
\text{RR}^r_i = \frac{P(M|G,i)}{P(M|G^C,i)} = \frac{a_i/(a_i + b_i)} {c_i/(c_i + d_i)}
$$
so that $P(M|G,i) = \text{RR}^r_i P(M|G^C,i)$, implying that the chances an article in G  is mentioned are $\text{RR}^r_i$ times the chances an article not in G is mentioned.
%. If $ \hat \psi_{i}^R >1$, then the chances of being mentioned are  $100 ( \hat \psi_{i}^R - 1)\%$ higher for articles in G compared to articles not in G. The interpretation is the same for $\hat{\psi}^R_{MH}$ but refers to all strata rather than a specific strata.

The column risk ratio is
$$
\text{RR}^c_i = \frac{P(G|M,i)}{P(G|M^C,i)} = \frac {a_i/(a_i + c_i)} {b_i/(b_i + d_i)}
$$
so that $P(G|M,i) = \text{RR}^c_i P(G|M^C,i)$, implying that the chances a mentioned article is in G are $\text{RR}^c_i$ times the chances a not mentioned article is in G.  %. If $ \hat \psi_{i}^C  >1$, then the chances of being in G are  $100 ( \hat \psi_{i}^C  - 1)\%$ higher for articles that are mentioned compared to articles that are not mentioned.  The interpretation is the same for MHq and $\hat{\psi}^C_{MH}$ but refers to the whole population rather than a specific strata.

For a comparison to the world, let $W= G \cup G^C$.  Recall that MHRR approximates the row risk ratio.  Since $P(G|i)$ is the fraction of the world's articles that belong to group G, we write $f_i = P(G|i)= \frac{a_i+b_i}{n_i}$.  The comparison is
 \begin{align}
 \frac{P(M|G,i)}{P(M|W,i)} = &  \frac{P(M|G,i)}{P(M|i)}\label{eq:worldcomprow}\\
= &  \frac {a_i/ (a_i+b_i)} {(a_i + c_i)/n_i} \nonumber\\
= & \frac {\frac{a_i/(a_i + b_i)} {c_i/(c_i + d_i)}  }{ 1 + \frac{a_i+b_i}{n_i} (\frac{a_i/(a_i + b_i)} {c_i/(c_i + d_i)} -1) } \nonumber\\
= & \frac{\text{RR}^r_i}{1+ f_i (\text{RR}^r_i -1)}. \nonumber
\end{align}
The calculation can also be done purely in probability using the law of total probability. Note that $\text{RR}^r_i >1$ implies $ 1<\frac{P(M|G,i)}{P(M|W,i)} < \text{RR}^r_i$.

%
%\[ \hat \gamma_i =  \frac {a_i/ (a_i+b_i)} {(a_i + c_i)/n_i}  =  \frac{P(M|G,i)}{P(M,i)} \]
%\[ P(M,i) \hat \gamma_i  = P(M|G,i) \]
%or the chances an article in G  is mentioned is $\hat \gamma_{i}$ times the chances an article (in the world) is mentioned.
%%. If $ \hat  \gamma_i >1$, then the chances of being mentioned are  $100 ( \hat  \gamma_i - 1)\%$ higher for articles in G compared to articles in the world.

Recall that MHq approximates the column risk ratio.  On each strata it is $\text{RR}^c_i = \frac{P(G|M,i)}{P(G|M^C,i)}$.
A direct comparison to the world is $\frac{P(G|M,i)}{P(W|M,i)}$, but
\begin{equation}
\label{eq:worldcompcolumn}
\frac{P(G|M,i)}{P(W|M,i)} < \text{RR}^c_i
\end{equation}
because $P(W|M,i) = 1$.

\vfill
\pagebreak

\section{ }
\label{app:Var[Estimators]}

\begin{center}
\textbf{Variance of ln(MHq) }
\end{center}

In this appendix, we derive formula (\ref {varestimator2}).  We use notation given in the first paragraph of Appendix \ref{app:convergence}.  In addition, let $\theta=\ln(\psi)$, $R_i = \frac{A_i n_{2i}}{A_i+B_i+n_i}$, and $S_i =\frac{B_in_{1i}}{A_i+B_i+n_i}$.  Noting that $a_i + c_i= n_{1i}$ and $b_i+d_i = n_{2i}$, these expressions of random variables correspond to the table entry definitions of $R_i$ and $S_i$ given above Eq. (\ref {varestimator2}). Let $R =\sum_{i=1}^K R_i$, $S =\sum_{i=1}^K S_i$, and $\theta = \ln(\psi) = \ln(\frac RS)$.

\vspace{0.2 cm}

\noindent

\subsection{An estimator in terms of expected values and variances of $R_i$ and $S_i$\nopunct}
\hfill\quad

An estimator for the variance of $\hat \theta =\ln (\text{MHq})$ is:
\begin{align}
\text {Var} [\hat \theta] & \approx \frac1{E[R]E[S]} \sum_{i=1}^K  ( \frac 1{\psi_i} (\text {Var}[R_i ] + E[R_i ]^2 ) \label{varestimator1rv} \\
&\qquad - 2 (\text {Cov} [R_i,S_i] + E[R_i]E[S_i]) + \psi_i(\text {Var} [S_i] + E[S_i] ^2 )    ). \nonumber
 \end{align}

\noindent
\textit{Derivation.}
To obtain the estimator (\ref{varestimator1rv}), proceed as follows.  Begin with the estimate
\[ \displaystyle \text {Var} [\hat \theta] \approx \frac { {\text {Var}}[R- \psi S]}{E[R]^2}.\]
The derivation is analogous to the calculations for the Mantel-Haenszel odds ratio and risk ratio (Greenland \& Robins, 1985; Phillips \& Holland, 1985; Robins, Breslow, \& Greenland, 1986).  Following the reasoning in Greenland (1989),

\begin{align*}
\text {Var} [\hat \theta]\approx &  \frac{\sum_{i=1}^K \text {Var}[R_i- \psi S_i]}{E[R]^2} \\
\approx &  \frac{\sum_{i=1}^K E[(R_i- \psi S_i)(R_i- \psi S_i])]}{E[R]^2} \\
\approx & \frac{\psi \sum_{i=1}^K E[(\frac1{\psi}R_i- S_i)(R_i- \psi S_i])]} {E[R]^2} \\
\approx & \frac{\frac{E[R]}{E[S]} \sum_{i=1}^K E[(\frac1{\psi_i}R_i- S_i)(R_i- \psi_i S_i])]} {E[R]^2} \\
\approx & \frac{ \sum_{i=1}^K E[(\frac1{\psi_i}R_i- S_i)(R_i- \psi_i S_i])]} {E[R]E[S]} \\
\approx & \frac{ \sum_{i=1}^K (\frac1{\psi_i}E[R_i^2] - 2 E[S_iR_i] + \psi_i E[S_i^2])]} {E[R]E[S]}. \\
 \end{align*}
To obtain (\ref{varestimator1rv}) use
$E[R_i^2] = \text {Var} [R_i] +E[R_i]^2$, $E[S_i^2] = \text {Var} [S_i ] + E[S_i]^2$, and
$E[R_i S_i] = \text {Cov} [R_i,S_i] + E[R_i] E[S_i].$
\hfill QED

\subsection{Expected values and variances of $R_i$ and $S_i$\nopunct}
\hfill\quad

We next need estimators for $E[R_i], E[S_i], \text {Var}[R_i ], \text {Var}[S_i], \text { and } \text {Cov} [R_i,S_i]$.  These are derived below with the formulas for $\widehat{\text {Var}}[R_i]$, $\widehat{\text {Var}} [S_i]$, and $\widehat{\text {Cov} }[R_i,S_i]$ given in Eqs. (\ref{eq:varRi}), (\ref{eq:varSi}), and (\ref{eq:cov}).   The method is described in Chapter 6 of Wolter (2007).

Let $f(x,y)= \frac x{x+y+n}$, so $\frac{\partial f}{\partial x}= \frac {n+y}{(x+y+n)^2}$ and
$\frac{\partial f}{\partial y}=\frac {-x}{(x+y+n)^2}.$ The Taylor series for $f(x,y)$ centered at $(x_0,y_0)$ is

\begin{align}
\label{eq:TaylorRS}
f(x,y) = & \frac {x_0}{x_0+y_0+n_i} + \\
& \frac {n+y_0}{(x_0+y_0+n_i)^2} {(x-x_0)} -  \frac {x_0}{(x_0+y_0+n_i)^2}(y-y_0) + \text{higher order terms.}
\nonumber
\end{align}

Since $R_i = (n_{2i}) f( A_i,B_i)$, let  $(x_0,y_0)=(E[A_i],E[B_i])$ and take the variance,

\begin{equation}
\text{Var} [R_i] \approx
n_{2i}^2 \left( \frac {(n_i+E[B_i])^2}{(E[A_i]+E[B_i]+n_i)^4} \text{Var}[A_i] + \frac {E[A_i]^2}{(E[A_i]+E[B_i]+n_i)^4}\text{Var}[B_i]\right).
\end{equation}
Expressed in terms of the binomial parameters,
\begin{equation}
\text{Var} [R_i]\approx
n_{2i}^2 \left( \frac {(n_i+n_{2i}p_{2i})^2}{(n_{1i}p_{1i}+n_{2i}p_{2i}+n_i)^4} n_{1i}p_{1i}(1-p_{1i}) \right. \left.
+ \frac {(n_{1i}p_{1i})^2}{(n_{1i}p_{1i}+n_{2i}p_{2i}+n_i)^4}n_{2i}p_{2i}(1-p_{2i})\right) \\
\end{equation}
and the estimator in terms of the table entries is
\begin{equation}
\label{eq:varRi}
\widehat {\text{Var}}[R_i]=
(b_i+d_i)^2 \left( \frac {(n_i+b_i)^2}{(a_i+b_i+n_i)^4} \frac{a_ic_i }{a_i+c_i} +  \frac {a_i^2}{(a_i+b_i+n_i)^4}\frac{b_id_i }{b_i+d_i}\right).
\end{equation}

To compute the variance of  $S_i $, write $S_i = (n_{1i}) f( B_i,A_i)$ and take the Taylor series at   $(x_0,y_0)=(E[B_i],E[A_i])$.  The  variance expressed in terms of the binomial parameters is
\begin{equation}
\label{eq:varSiPar}
\text{Var} [S_i] \approx
 n_{1i}^2 \left(  \frac {(n_{2i}p_{2i})^2 n_{1i}p_{1i}(1-p_{1i})} {(n_{1i}p_{1i}+n_{2i}p_{2i}+n_i)^4}
   +  \frac {(n_i+n_{1i}p_{1i})^2 n_{2i}p_{2i}(1-p_{2i})}{(n_{1i}p_{1i}+n_{2i}p_{2i}+n_i)^4}
\right)
\end{equation}
and the estimator in terms of the table entries is
\begin{equation}
\label{eq:varSi}
\widehat {\text{Var}} [S_i]=
(a_i+c_i)^2 \left(  \frac {b_i^2}{(a_i+b_i+n_i)^4} \frac{a_ic_i }{a_i+c_i} +  \frac {(n_i+a_i)^2}{(a_i+b_i+n_i)^4}\frac{b_id_i }{b_i+d_i}
\right).
\end{equation}

To compute the covariance, we use 6.2.5 (Wolter, 2007),
\begin{align*}
\text {Cov} [R_i,S_i] =& \text {Cov} [ n_{2i} f( A_i,B_i), n_{1i} f( B_i,A_i)]\\
\approx& \text{Var} [A_i] \left.\frac{\partial}{\partial A_i} (n_{2i} f( A_i,B_i)) \right|_{( E[A_i],E[B_i] )}
\left.\frac{\partial}{\partial B_i}(n_{1i} f( B_i,A_i))\right|_{( E[B_i], E[A_i])} \\
&\quad
+ \text{Var} [B_i]   \left. \frac{\partial}{\partial B_i} f( A_i,B_i))\right|_{( E[A_i],E[B_i] )}
  \left. \frac{\partial}{\partial A_i}(n_{1i} f( B_i,A_i))\right|_{( E[B_i], E[A_i])}.
\end{align*}
Expressed in terms of the binomial parameters, we have
\begin{equation}
\label{eq:covpar}
 \text {Cov} [R_i,S_i] \approx  - \frac{n_{1i}^2 n_{2i}^2 p_{1i}p_{2i} }{(n_{1i}p_{1i}+n_{2i}p_{2i}+n_i)^4}
 \left( (1-p_{1i}) (n_{2i}p_{2i}+n_i)
 +(1-p_{2i}) (n_{1i}p_{1i}+n_i) \right),
\end{equation}
and the covariance estimator in terms of the data is
\begin{equation}
\label{eq:cov}
\widehat { \text {Cov}} [R_i,S_i] =  - \frac{(a_i+c_i) (b_i+d_i)}{(a_i+b_i+n_i)^4} \left(\frac{a_i b_i c_i (b_i+n_i)}{a_i+c_i}+\frac{a_i b_i d_i (a_i+n_i)}{b_i+d_i}\right).
\end{equation}

\subsection{Derivation of Formula (\ref{varestimator2})\nopunct}
\hfill\quad

To arrive at Formula (\ref{varestimator2}), start with Formula (\ref {varestimator1rv}) and replace $\psi_i$, the expected values, and variances with the appropriate estimates.  The strata column risk ratio is $\hat\psi_i = \frac{a_i(b_i+d_i)}{b_i(a_i+c_i)}$.  The expected values are
$\hat E[R_i]=\frac{a_i (b_i+d_i)}{a_i+b_i+n_i}$, $\hat E[S_i]=\frac{b_i( a_i+c_i)}{a_i+b_i+n_i}$, $\hat E[R]=\sum \frac{a_i (b_i+d_i)}{a_i+b_i+n_i}$, and $\hat E[S]=\sum \frac{b_i( a_i+c_i)}{a_i+b_i+n_i}$.
Finally, the formulas for $\widehat{\text {Var}}[R_i]$, $\widehat{\text {Var}} [S_i]$, and $\widehat{\text {Cov} }[R_i,S_i]$ are given in Eqs. (\ref{eq:varRi}), (\ref{eq:varSi}), and (\ref{eq:cov}).  This completes the derivation.

% \newpage

%%%%% END APPENDIX

\vfill
\pagebreak

\begin{center}
\textbf{References}
\end{center}

\hangindent=0.5 in
Agresti, A. (2019). \textit{An introduction to categorical data analysis} (3rd ed.) Hoboken, NJ, USA: Wiley.

\hangindent=0.5 in
Agresti, A. \& Hartzel, J. (2000).  Tutorial in biostatistics strategies for comparing treatments on a binary response
with multi-centre data.  \textit{Statistics in Medicine, 19}, 1115-1139.

\hangindent=0.5 in
Böhning, B., Sangnawakij, P., \& Holling, H. (2021). Confidence interval estimation for the Mantel–Haenszel estimator of the risk ratio and risk difference in rare event meta-analysis with
emphasis on the bootstrap. \textit{Journal of Statistical Computation and Simulation} DOI: 10.1080/00949655.2021.1991347.

\hangindent=0.5 in
Bornmann, L., \& Haunschild, R. (2018a). Normalization of zero-inflated data: An empirical analysis of a new indicator family and its use with altmetrics data. \textit{Journal of Informetrics, 12}(3), 998–1011.

\hangindent=0.5 in
Bornmann, L., \& Haunschild, R. (2018b). Do altmetrics correlate with the quality of papers? A large-scale empirical study based on F1000Prime data. \textit{ PloS One, 13}(5), 0197133. https://doi.org/10.1371/journal.pone.0197133.

\hangindent=0.5 in
Bornmann, L., \& Haunschild, R. (2018c). Alternative article-level metrics: The use of alternative metrics in research evaluation. \textit{ EMBO Reports, 19}(12), e47260. DOI: https://doi.org/10.15252/embr.201847260

\hangindent=0.5 in
Bornmann L.,  \& Haunschild R. (2019) Societal impact measurement of research papers. In: Glänzel W., Moed H.F., Schmoch U., Thelwall M. (eds) \textit{ Springer handbook of science and technology indicators.} Springer Handbooks. Springer.\\
https://doi.org/10.1007/978-3-030-02511-3\_23

\hangindent=0.5 in
Bornmann, L., Haunschild, R.,  \& Adams, J. (2018) Convergent validity of altmetrics and case studies for assessing societal impact: an analysis based on UK Research Excellence Framework (REF) data.  In Costas, R., Franssen, T.,   \& Yegros-Yegros, A. (Eds.), \textit{ STI 2018 Conference proceedings: Proceedings of the 23rd International Conference on Science and Technology Indicators}. http://hdl.handle.net/1887/65305

\hangindent=0.5 in
Bornmann, L., Haunschild, R.,  \& Adams, J. (2019) Do altmetrics assess societal impact in a comparable way to case studies? An empirical test of the convergent validity of altmetrics based on data from the UK research excellence framework (REF).  \textit{ Journal of Informetrics, 13}(1), 325-340.

\hangindent=0.5 in
Bornmann, L., Haunschild, R., \& Mutz, R. (2018). MHq indicators for zero-inflated count data—A response to Smolinsky and Marx (2018). \textit{Journal of Informetrics, 12}(3), 1012–1014. http://dx.doi.org/10.1016/j.joi.2018.08.001

\hangindent=0.5 in
Bornmann, L., Haunschild, R., \& Mutz, R. (2019). MHq indicators for zero-inflated count data—A response to the comment by Smolinsky (in press).  \textit{Journal of Informetrics 13}, 464–465.  https://doi.org/10.1016/j.joi.2019.02.004

\hangindent=0.5 in
Copiello, S. (2020a). Multi-criteria altmetric scores are likely to be redundant with respect to a subset of the underlying information. \textit{ Scientometrics 124}, 819–824. https://doi.org/10.1007/s11192-020-03491-9

\hangindent=0.5 in
Copiello, S. (2020b).  Other than detecting impact in advance, alternative metrics could act as early warning signs of retractions: tentative findings of a study into the papers retracted by \textit{PLoS ONE}. \textit{ Scientometrics 125}, 2449–2469. https://doi.org/10.1007/s11192-020-03698-w

\hangindent=0.5 in
Davis, L.D. (1985) Generalization of the Mantel-Haenszel Estimator to Nonconstant Odds Ratios.  \textit{Biometrics, 41}(2), 487-495.

\hangindent=0.5 in
Fleiss, J., Levin, B., \& Paik, M. C. (2003). \textit{Statistical methods for rates and proportions} (3rd ed.). Hoboken, NJ, USA: Wiley.

\hangindent=0.5 in
Greenland, S., (1989).  Generalized Mantel-Haenszel Estimators for K 2 x J tables. \textit{Biometrics, 45}(1),183-191.

\hangindent=0.5 in
Greenland, S. \&  Robins, J.M. (1985).  Estimation of a common effect parameter from sparse follow-up data.  \textit{Biometrics 41}, 55-68.

\hangindent=0.5 in
Haunschild, R., Leydesdorff, L., Bornmann, L., Hellsten, I.,  \& Marx, W. (2019). Does the public discuss other topics on climate change than researchers? A comparison of explorative networks based on author keywords and hashtags, \textit{ Journal of Informetrics, 13}(2), 695-707.

\hangindent=0.5 in
Haustein, S., Sugimoto, C., and Larivi\`ere, V. (2015). Guest editorial: social media in scholarly communication. \textit{ Aslib Journal of Information Management, 67}(3).
http://doi.org/10.1108/AJIM-03-2015-0047.

\hangindent=0.5 in
Jiangbo, L., Liang, Z.,  \& Chunlin, J. (2020) Research on influence evaluation of humanities and social sciences academic monographs from the perspective of altmetrics: Comparative analysis based on BkCI, Amazon, and Goodreads. \textit{ Journal of the China Society for Scientific and Technical Information, 39}(9), 896-905.

\hangindent=0.5 in
Kassab, O., (2019) Does public outreach impede research performance? Exploring the ‘researcher’s dilemma’ in a sustainability research center, \textit{ Science and Public Policy, 46}(5), 710–720, https://doi.org/10.1093/scipol/scz024

\hangindent=0.5 in
Kassab, O., Bornmann, L.,  \& Haunschild, R. (2020) Can altmetrics reflect societal impact considerations?: Exploring the potential of altmetrics in the context of a sustainability science research center. \textit{ Quantitative Science Studies,1}(2), 792-809.

\hangindent=0.5 in
Katz, D., Baptista, J., Azen, S.P.,  \& Pike, M.C. (1978). Obtaining confidence intervals for the risk ratio in cohort studies. \textit{ Biometrics, 34}(3), 469-474.

Klingenberg, B. (2014). A new and improved confidence interval for the Mantel–Haenszel risk difference. \textit{Statistics in Medicine, 33}(17), 2968-2983.

\hangindent=0.5 in
Kudryavtsev, L.D.,(1989). Double limit. In Hazewinkel, M. (Ed), \textit{Encyclopaedia of mathematics} (Volume 3, p. 299). Springer.

\hangindent=0.5 in
Lachin, J. M. (2009). \textit{ Biostatistical methods: The assessment of relative risks} (2nd ed.). Wiley-Interscience.

\hangindent=0.5 in
National Research Council 2012. Improving measures of science, technology, and innovation: Interim report. Washington, DC: The National Academies Press. \\
https://doi.org/10.17226/13358.

Noma, H., \& Nagashima, K. (2016). A note on the Mantel-Haenszel Estimators when the common effect assumptions are violated. \textit{Epidemilogical Methods, 5}(1), 19-35.

\hangindent=0.5 in
Ortega, J.L. (2020). Altmetrics data providers: A meta-analysis review of the coverage of metrics and
publications. \textit{ El profesional de la informaci\'on, 29}(1), e290107. https://doi.org/10.3145/epi.2020.ene.07

\hangindent=0.5 in
Phillips, A. and Holland, P.W. (1987). Estimators of the variance of the Mantel-Haenszel log-odds-ratio estimate.  \textit{Biometrics, 43}(2), 425-431.

\hangindent=0.5 in
Robins, J., Breslow, N. E., \& Greenland, S. (1986). Estimators of the Mantel-Haenszel variance consistent in both sparse data and large-strata limiting models. \textit{Biometrics 42}, 311-324.

\hangindent=0.5 in
Smolinsky, L.  (2019). Odds ratios and Mantel-Haenszel quotients. \textit{ Journal of Informetrics 13}, 462–463.

\hangindent=0.5 in
Smolinsky, L. \& Marx, B.D. (2018). Odds ratios, risk ratios, and Bornmann and Haunschild’s new indicators. \textit{ Journal of Informetrics 12},  732–735.

\hangindent=0.5 in
Tahamtan, I.  \& Bornmann, L. (2020). Altmetrics and societal impact measurements: Match or mismatch?  A literature review. \textit{ El profesional de la informaci\'on, 29}(1), e290102. https://doi.org/10.3145/epi.2020.ene.02

\hangindent=0.5 in
Thelwall, M. (2017a). Three practical field normalised alternative indicator formulae for research evaluation.  \textit{ Journal of Informetrics, 11}(1), 128–151.  http://dx.doi.org/10.1016/j.joi.2016.12.002

\hangindent=0.5 in
Thelwall, M. (2017b). \textit{ Web indicators for research evaluation: A practical guide}.  Morgan \& Claypool.

\hangindent=0.5 in
Universities and Higher Education in Europe. (2020). \textit{uniRank}. https://www.4icu.org/Europe/

\hangindent=0.5 in
Wang, X., Lv, T.  \& Hamerly, D. (2019), How do altmetric sources evaluate scientific collaboration? An empirical investigation for Chinese collaboration publications, \textit{ Library Hi Tech, 38}(3) 563-576. https://doi.org/10.1108/LHT-05-2019-0101

\hangindent=0.5 in
Williams, D. (1991). \textit{ Probability with Martingales}. Cambridge University Press.
ISBN 10:1139642987

\hangindent=0.5 in
Wolter, K.M. (2007).  \textit{Introduction to variance estimation} (2nd ed.). Springer.

\end{document}